\documentclass[draftcls,12pt,onecolumn]{IEEEtran}
\usepackage[latin9]{inputenc}
\usepackage{color}
\usepackage{array}
\usepackage{mathtools}
\usepackage{multirow}
\usepackage{amsmath}
\usepackage{amsthm}
\usepackage{amssymb}

\makeatletter

\providecommand{\tabularnewline}{\\}

\theoremstyle{plain}
\newtheorem{thm}{\protect\theoremname}
\theoremstyle{definition}
\newtheorem{defn}[thm]{\protect\definitionname}
\theoremstyle{definition}
\newtheorem{example}[thm]{\protect\examplename}
\theoremstyle{plain}
\newtheorem{cor}[thm]{\protect\corollaryname}

\usepackage{amsfonts}
\usepackage{algorithmic}
\usepackage{algorithm}
\usepackage{cite}
\usepackage{xcolor}

\makeatother

\providecommand{\corollaryname}{Corollary}
\providecommand{\definitionname}{Definition}
\providecommand{\examplename}{Example}
\providecommand{\theoremname}{Theorem}

\begin{document}

\title{CASO: Cost-Aware Secure Outsourcing of General Computational Problems}

\author{Kai Zhou and Jian Ren\thanks{The authors are with the Department of Electrical and Computer Engineering,
Michigan State University, East Lansing, MI 48824-1226, Email: \{zhoukai,
renjian\}@msu. edu}}
\date{October 19, 2015}
\maketitle
\begin{abstract}
Computation outsourcing is an integral part of cloud computing.  It
enables end-users to outsource their computational tasks to the cloud
and utilize the shared cloud resources in a pay-per-use manner.  However,
once the tasks are outsourced, the end-users will lose control of
their data, which may result in severe security issues especially
when the data is sensitive.  To address this problem, secure outsourcing
mechanisms have been proposed to ensure security of the end-users'
outsourced data.  In this paper, we investigate outsourcing of general
computational problems which constitute the mathematical basics for
problems emerged from various fields such as engineering and finance.
 To be specific, we propose affine mapping based schemes for the problem
transformation and outsourcing so that the cloud is unable to learn
any key information from the transformed problem.  Meanwhile, the
overhead for the transformation is limited to an acceptable level
compared to the computational savings introduced by the outsourcing
itself.  Furthermore, we develop cost-aware schemes to balance the
trade-offs between end-users' various security demands and computational
overhead.  We also propose a verification scheme to ensure that the
end-users will always receive a valid solution from the cloud.  Our
extensive complexity and security analysis show that our proposed
Cost-Aware Secure Outsourcing (CASO) scheme is both practical and
effective.  \end{abstract}


\section{Introduction}

Cloud computing paradigm provides end-users an on-demand access to
a shared pool of computing resources, such as computational power
and storage.  It enables the end-users to utilize those resources
in a pay-per-use manner instead of purchasing expensive equipment
upfront.  Computation outsourcing is a key component of cloud computing.
 It enables the resource-constrained end-users to outsource their
computational tasks to the cloud servers.  Then the tasks are processed
in the cloud servers and solutions are returned to the end-users.
 The technical and economic advantages make computation outsourcing
a promising application for cloud computing. 

However, security has become one of the major concerns that prevent
computation outsourcing from being widely adopted.  When the end-users
outsource their tasks to the cloud, they inevitably lose control of
their own data, while the cloud servers will get full access to not
only the problem itself but also the input, the intermediate computational
results and the output of the problem, which may contain sensitive
end-user data, such as financial statistics or health records.  As
a result, the end-users' privacy is totally exposed to the cloud.
 Furthermore, the cloud may have the motivation to cheat in the computation
process thus false solutions may be returned to the end-users.  This
is because the computing resources are regarded as a kind of commodity
and the cloud may try to reduce the cost by simply not investing enough
computing resources as it has claimed.  For example, the cloud may
just return a trivial result for an outsourced task thus saving a
lot of resources.  All these issues call for designs of more secure
and privacy-preserving outsourcing mechanisms that can also provide
end-users the ability to validate the received results. 

To address the aforementioned issues, researchers have proposed various
secure outsourcing schemes for different types of computational problems,
such as sequence comparison \cite{atallah2005secure,blanton2012secure,blanton2010secure},
linear algebra\cite{atallah2010,benjamin2008private,wang2011infocom,seitkulov2013new}
and modular exponentiation\cite{hohenberger2005securely,chen2012new}.
 The techniques utilized by these schemes can be divided into two
categories: encryption based schemes and disguising based schemes.
 Researchers from the cryptography community are trying to develop
specific encryption schemes under which computation can be carried
out on encrypted data.  For instance, in \cite{gentry2009} the authors
proposed a fully homomorphic encryption scheme under which an arbitrary
boolean circuit can be evaluated directly over the encrypted data.
 Based on this homomorphic encryption and Yao's garbled circuit \cite{yao1982},
the authors in \cite{gennaro2010} designed a secure outsourcing scheme
for arbitrary functions where the input and output privacy are protected
and the results can be verified in a non-interactive way.  However,
the main drawback of this type of schemes is that they all require
expensive encryption operations thus making it impractical to be carried
out in the cloud scenario.  Researchers in the theoretic computer
science community have developed some disguising techniques to transform
different types of computational problems to disguised forms so that
the private information of the original problems is concealed.  Based
on disguising, the authors in \cite{atallah2002} and \cite{atallah2010}
developed schemes to securely outsource some basic scientific operations
such as matrix multiplication, matrix inversion and convolution.  More
recently, secure and practical outsourcing schemes were proposed in
\cite{wang2011infocom}\cite{znie2014efficient} for linear programming.
 In \cite{wang2011icdcs}\cite{chen2015new}, the authors focused
on outsourcing of large-scale systems of linear equations.  However,
the above mentioned disguising techniques are specially designed for
a particular kind of scientific computation, mostly lies in the scope
of linear algebra.  Thus the application of the proposed schemes is
quite limited. 

In this paper, we aim at developing a secure outsourcing scheme that
is suitable for general computational problems.  The challenges come
from various aspects.  First, we target at general computational problems
which cover the scope of linear and non-linear problems such as system
of equations (linear or non-linear), linear programming and convex
optimization.  Due to the different natures of these problems, it
is extremely challenging to design an outsourcing scheme suitable
for various kinds of computational problems.  Second, in the cloud
scenario, the end-users are resource-constrained which means that
the operations can be implemented before and after the outsourcing
are quite limited.  Third, the end-users vary from handheld mobile
devices to desktop workstations in terms of resource constraints and
security requirements.  Thus it is not easy to design a scheme that
can meet the requirements of various end-users.  Finally, our preliminary
investigation shows that a more complex pre-processing of the problem
will ensure a more secure outsourcing process.  However, it also creates
more computational burden on the end-users.  Thus there exists a trade-off
between the computational complexity that the end-users can afford
and the security they can get in return.  All these concerns make
it extremely hard to design a secure outsourcing scheme for general
computational problems. 

To deal with the aforementioned challenges, we propose a secure outsourcing
scheme based on affine mappings.  The basic idea is that before outsourcing,
the independent variables of the computational problem is mapped to
a new group of variables through an affine mapping.  Correspondingly,
the original problem is transformed to a new form that can be securely
outsourced to the cloud.  Then the cloud can generate valid results
from the transformed problem and return the results of the transformed
problem back to the end-user.  By applying an inverse affine transformation
on the results returned from the cloud, the end-user can derive the
valid results to the original problem efficiently at the local environment.
 We prove that the proposed outsourcing scheme can ensure security
of the private information of the original problem.  

The contributions of this paper can be summarized as follows: 
\begin{itemize}
\item We propose a cost-aware secure outsourcing scheme (CASO) that is generally
suitable for a wide variety of computational problems, such as system
of equations, linear programming and convex optimization. 
\item We investigate the trade-off between the computational complexity
and security such that end-users can choose the most suitable outsourcing
scheme according to their own resource constraints and security demands.
 
\item Our analysis and performance comparison demonstrate that CASO is much
more efficient than the existing schemes with comparable security
levels.  
\item We also introduce a verification process which enables the end-users
to verify the validity of the results returned from the cloud servers.
 
\end{itemize}
The rest of this paper is organized as follows.  In Section \ref{sec:Problem-Statement},
we introduce our system model, threat model and our design goals.
 In Section \ref{sec:Secure-Outsourcing-Based}, we present the basic
idea of CASO based on affine mappings.  We use system of linear equations
as a case study to illustrate our cost-aware design philosophies in
Section \ref{Sec:Cost-Aware-Design}.  We extend our design to non-linear
problems in Section \ref{sec:Extension-to-Non-linear} and the result
verification scheme is introduced in Section \ref{Sec:Results-Verification}.
 We evaluate the performance of our scheme by comparing it with several
existing works and giving some numeric results in Section \ref{sec:Evaluation}.
 We conclude our work in Section \ref{sec:Conclusion}.

\section{Problem Statement\label{sec:Problem-Statement}}

\subsection{System and Threat Model}

We consider a system consisting of two entities: the end-user and
the cloud. Suppose that an end-user wants to solve a general computational
problem denoted by $F(\mathbf{x})$, where $\mathbf{x}=(x_{1},x_{2},\cdots,x_{n})$
is a series of independent variables. Note that $F(\mathbf{x})$ describes
a general computational problem not necessarily restricted to a function.
For example, it can be a system of equations or an optimization problem.
However, due to lack of resources, the end-user needs to outsource
the problem to the cloud which is considered to have infinite computing
resources. Before outsourcing, the end-user will transform the original
problem at the local side in order to prevent information leakage.
On receiving the transformed problem, the cloud server will carry
out the computing process and return the solution to the end-user.
Then at the local side, an inverse transformation is carried out on
the solution returned from the cloud to recover the solution of the
original problem. Based on the transformation and the information
returned by the cloud, the end-user is able to verify the validity
of the received solution. 

As the problem is outsourced to the cloud, the end-users totally lose
control of their own data. The private data and the computational
result will be revealed to the cloud. There are at least three reasons
that the cloud cannot be fully trusted. First, the cloud could be
honest but curious. That is the cloud may collect any information
that could be revealed by the problem. Second, the cloud is profit-motivated.
As the end-users pay for the resources during the computation process,
the cloud may reduce the cost by utilizing less resources and simply
returning some trivial results. Third, cloud is a shared environment,
it is hard to secure individual data using just regular processor.
As a result, a secure outsourcing scheme should not only prevent the
private information from being exposed to the cloud but also guarantee
that the end-users can receive valid results. As a result, we have
to develop effective mechanisms to protect this private end-user information.

\subsection{Design Goals}

Under the above system and threat model, our proposed outsourcing
scheme should achieve the following goals:
\begin{enumerate}
\item \textbf{Soundness}: Given that the cloud is trustworthy, the transformation
on the problem and the inverse transformation of the returned result
should guarantee that the recovered solution is correct. 
\item \textbf{Security}: When the problem is outsourced to the cloud, it
should be computationally infeasible for the cloud server to infer
the coefficient matrix, the input and output of the original outsourced
problem. 
\item \textbf{Verifiability}: In case that the cloud cannot be fully trusted,
the end user should have the ability to verify the validity of the
solution returned by the cloud. 
\item \textbf{Efficiency}: The outsourcing scheme should be efficient in
computation and communication. For computation, the overhead caused
by the problem transformation, the inverse transformation and the
result verification should be limited to $\mathcal{O}(n^{2})$. For
communication, the overhead caused by the outsourcing process should
be in the same level as that of outsourcing the original problem. 
\item \textbf{Cost-Awareness}: The end-users can select different outsourcing
strategies according to their own computational constraints and security
demands in a cost-aware manner. 
\end{enumerate}

\section{Secure Outsourcing Based on Affine Mapping\label{sec:Secure-Outsourcing-Based}}

In this section, we will first present the basic framework of the
proposed CASO. Then we introduce the equivalence concept of two general
computational problems. Under this definition, we explain in detail
the process of problem transformation. At the end of this section,
we prove the soundness of the CASO.

\subsection{Basic Framework}

As mentioned previously, we assume that the end user has a general
computational problem $F(\mathbf{x})$ to be solved. Due to the lack
of resources, the end user needs to outsource $F(\mathbf{x})$ to
the cloud. We formally divide the outsourcing process into the following
phases. 
\begin{enumerate}
\item \textbf{Problem Transformation}: \emph{$\mathsf{ProbTran}$} $\{\mathbf{S},F(\mathbf{x})\}\rightarrow\{G(\mathbf{y})\}$.
In this phase, the end user first generates a key $\mathbf{S}$ which
is kept secret at the local side during the whole process. Based on
this secret key, the end user transforms $F(\mathbf{x})$ to a new
form $G(\mathbf{y})$, where $\mathbf{y}$ is the new input. 
\item \textbf{Cloud Computation}: \emph{$\mathsf{CloudCom}$} $\{G(\mathbf{y})\}\to\{\mathbf{y}^{*},\Phi\}$.
On receiving the transformed problem $G(\mathbf{y})$, the cloud carries
out the necessary computation and gives the solution $\mathbf{y^{*}}$
as well as a proof $\Phi$ of the validity of the returned solution. 
\item \textbf{Result Recovery}\textbf{\textcolor{black}{{} and Verification}}:
\emph{$\mathsf{RecVeri}$} $\{\mathbf{y}^{*},\mathbf{S},\Phi\}\to\{\mathbf{x}^{*},\Lambda\}$.
By utilizing the secret key $\mathbf{S}$, the end-user recovers solution
$\mathbf{x^{*}}$ to the original problem from $\mathbf{y^{*}}$.
Based on the proof $\Phi$, the end-user gives the decision $\Lambda=\{\mathsf{Ture},\mathsf{False}\}$,
indicating the validity of $\mathbf{x^{*}}$. 
\end{enumerate}

\subsection{Problem Transformation\label{sub:Problem-Transformation}}

The basic idea of problem transformation is to map the independent
variables of the problem to a new group of variables such that the
original problem is transformed to a new form.  To be specific, suppose
the original problem is $F(x)$.  We assume that $\psi:\mathbb{R}^{n}\rightarrow\mathbb{R}^{n}$
is a general one-to-one mapping function.  Let $\mathbf{x}=\psi(\mathbf{y})$,
then $F(\mathbf{x})=F(\psi(\mathbf{y}))=(F\circ\psi)(\mathbf{y})=G(\mathbf{y})$.
 In this way, the original input $\mathbf{x}$ can be transformed
to input $\mathbf{y}$ with the relationship determined by the function
$\psi$.  Below, we give the equivalence definition of two computational
problems. 
\begin{defn}[Equivalence]
\label{Def:Equivalence} Denote a set of computational problems as
$\Omega=\{\Gamma\;|~\Gamma:\mathbb{R}^{n}\rightarrow\mathbb{R}^{n}\}$.
 For any $F\in\Omega$, if there exists a one-to-one mapping $\psi:\mathbb{R}^{n}\rightarrow\mathbb{R}^{n}$
such that $F(\mathbf{x})=F(\psi(\mathbf{y}))=(F\circ\psi)(\mathbf{y})=G(\mathbf{y})$,
then $F$ is said to be equivalent to $G$.  We denote it as $F\sim G$.
 The equivalent class of $F$ is denoted as $[F]=\{\Gamma\in\Omega\;|~\Gamma\sim F\}$.
 \end{defn}
\begin{thm}
The equivalence relation defined in Definition \ref{Def:Equivalence}
is well-defined.  \end{thm}
\begin{IEEEproof}
We only need to prove that the relation defined in Definition \ref{Def:Equivalence}
is reflexive, symmetric and transitive.  First, it is obvious that
for every $F\in\Omega$, if we select the one-to-one mapping $\psi$
to be the identity mapping, then we have $F(\mathbf{x})=F(\psi(\mathbf{y}))=F(\mathbf{y})$.
 Thus for every $F\in\Omega$, we have $F\sim F$ which demonstrates
the property of reflexivity.  Second, for $F,G\in\Omega$, if $F\sim G$,
then there exists a one-to-one mapping $\psi$ such that $F(\mathbf{x})=F(\psi(\mathbf{y}))=(F\circ\psi)(\mathbf{y})=G(\mathbf{y})$,
which indicates the existence of an inverse mapping $\psi^{-1}$ such
that $G(\mathbf{y})=(F\circ\psi)(\psi^{-1}(\mathbf{x}))=F(\mathbf{x})$.
 Thus we have $G\sim F$ and the property of symmetry holds.  To prove
the property of transitivity, assume that $F,G,H\in\Omega$ such that
$F\sim G$ and $G\sim H$.  This means that there are two one-to-one
mappings $\psi$ and $\phi$ such that $\mathbf{x}=\psi(\mathbf{y})$,
$F(\mathbf{x})=F(\psi(\mathbf{y}))=G(\mathbf{y})$ and $\mathbf{y}=\phi(\mathbf{z})$,
$G(\mathbf{y})=G(\phi(\mathbf{z}))=H(\mathbf{z})$.  Therefore, we
have $F(\mathbf{x})=F(\psi(\mathbf{y}))=F((\psi\circ\phi)(\mathbf{z}))=H(\mathbf{z})$.
 Since $\psi$ and $\phi$ are both one-to-one mappings, the mapping
$\psi\circ\phi$ is also one-to-one.  Thus from the definition we
have $F\sim H$ and the equivalence relation is transitive.  
\end{IEEEproof}
The following example illustrates the basic idea of problem transformation. 
\begin{example}
Suppose $F(\mathbf{x})$ represents a system of linear equations with
two independent variables: 
\begin{equation}
F(\mathbf{x}):=\left\{ \begin{array}{c}
x_{1}+2x_{2}=6\\
3x_{1}+x_{2}=3.
\end{array}\right.
\end{equation}
If we take the mapping function as $\mathbf{x}=(x_{1},x_{2})=\psi(\mathbf{y})=(2y_{2}+1,3y_{1}+2)$,
the original problem is transformed to 
\begin{equation}
G(\mathbf{y})=F(\psi(\mathbf{y}))=\left\{ \begin{array}{c}
6y_{1}+2y_{2}=1\\
3y_{1}+6y_{2}=-2.
\end{array}\right.
\end{equation}

Since the above mapping $\psi$ is one-to-one, we have $F\sim G$
according to Definition \ref{Def:Equivalence}. In fact, the solutions
to the two systems also satisfy the same mapping function. It is easy
to obtain the solutions as $\mathbf{y}^{*}=(\frac{1}{3},-\frac{1}{2})$
and $\mathbf{x}^{*}=(0,3)$, which satisfy $\mathbf{x}^{*}=\psi(\mathbf{y}^{*})$. 
\end{example}
The above example gives an insight of CASO. Based on a one-to-one
mapping $\psi$, the end-user first transforms the original problem
$F(\mathbf{x})$ to an equivalent form $G(\mathbf{y})$ that can be
securely outsourced to the cloud. Since the solutions to the two problem
satisfy $\mathbf{x}^{*}=\psi(\mathbf{y}^{*})$, the end-user can always
recover $\mathbf{x}^{*}$ from $\mathbf{y}^{*}$ returned by the cloud.
Thus the essence of our proposed scheme lies in finding a proper one-to-one
mapping that satisfies the various design goals. 
\begin{defn}
An affine mapping $\psi:\mathbb{R}^{n}\rightarrow\mathbb{R}^{n}$
is defined as a mapping from $\mathbf{x}\in\mathbb{R}^{n}$ to $\mathbf{y}\in\mathbb{R}^{n}$
satisfying $\mathbf{x}=\mathbf{Ky}+\mathbf{r}$, where $\mathbf{K}\in\mathbb{R}^{n\times n}$
is nonsingular and $\mathbf{r}\in\mathbb{R}^{n}$. 
\end{defn}
It is clear that as long as $\mathbf{K}$ is nonsingular, the affine
mapping defined above is a one-to-one mapping. The soundness of our
proposed scheme based on affine mapping is guaranteed by the following
theorem. 
\begin{thm}[Soundness]
Under the affine mapping, the transformed problem is equivalent to
the original problem. That is the end-user is guaranteed to be able
to recover the valid solution of the original problem from the solution
returned by the cloud. \end{thm}
\begin{IEEEproof}
The proof of soundness follows the definition of equivalence. The
affine mapping $\mathbf{x}=\mathbf{Ky}+\mathbf{r}$ is one-to-one
as long as $\mathbf{K}$ is non-singular. Thus by definition, $F\sim G$
under this affine mapping. Since the solutions to the two problems
satisfy $\mathbf{x}^{*}=\mathbf{K}\mathbf{y}^{*}+\mathbf{r}$, given
$\mathbf{y}^{*}$ returned by the cloud, the end-user is able to recover
$\mathbf{x}^{*}$ at the local side. 
\end{IEEEproof}
In the rest of this paper, we will show that our proposed affine mapping
based CASO will not only meet the end-user's resource constraints
but also conceal the end-user's private information. Furthermore,
CASO can provide a cost-aware trade-off so that the end-user can achieve
the desired security levels by selecting different outsourcing strategies
with different computational and communication overhead. In the following
analysis, we divide the computational problems into two categories:
linear systems and non-linear systems due to their different mathematical
properties.

\section{Cost-Aware Design for Linear Systems\label{Sec:Cost-Aware-Design}}

In this section, we present our cost-aware secure outsourcing scheme
for general computational problems. In the region of linear computation,
we deploy system of linear equations as a case study to show the principles
of our design. Then we show that the proposed CASO can be well extended
to linear programming.

\subsection{Outsourcing Scheme}

In the problem transformation phase, the end-user first generates
a one-time secret key $\mathbf{S}=\{\mathbf{K},\mathbf{r}\}$, where
$\mathbf{K}\in\mathbb{R}^{n\times n}$ is a non-singular matrix and
$\mathbf{r}\in\mathbb{R}^{n}$. Then $\mathbf{x}=\mathbf{Ky}+\mathbf{r}$
is a one-to-one mapping from $\mathbf{x}$ to $\mathbf{y}$. 

Suppose the computational problem is a system of linear equations
$\mathbf{A}\mathbf{x}=\mathbf{b}$, where $\mathbf{x,\mathbf{b}}\in\mathbb{R}^{n}$
and $\mathbf{A}$ is an $n\times n$ nonsingular matrix. The function
\emph{$\mathsf{ProbTran}$} $\{\mathbf{S},F(\mathbf{x})\}\to\{G(\mathbf{y})\}$
takes the secret key $\mathbf{S}=\{\mathbf{K},\mathbf{r}\}$ and the
linear system as input and generates the output as $\mathbf{AKy}=\mathbf{b-Ar}$.
Denote $\mathbf{A}'=\mathbf{AK}$ and $\mathbf{b}'=\mathbf{b-Ar}$
and the system is transformed to $G(\mathbf{y}):\mathbf{A}^{\prime}\mathbf{y}=\mathbf{b}'$
which can be outsourced to the cloud. 

In the phase of cloud computation, the cloud solves $G(\mathbf{y})$
utilizing the typical methods and returns the solution $\mathbf{y}^{*}$
to the end-user. Then in the result recovery phase, the end-user recovers
the solution to the original system of linear equations as $\mathbf{x}^{*}=\mathbf{K}\mathbf{y}^{*}+\mathbf{r}$.
The result verification will be discussed in detail in Section \ref{Sec:Results-Verification}.

\subsection{Design Analysis}

From the above outsourcing scheme, we can see that the computational
overhead for the end-user incurs both in the problem transformation
and the result recovery phase. To be more specific, in the problem
transformation phase, the end-user needs to calculate $\mathbf{AK}$
and $\mathbf{Ar}$. To recover the original solution $\mathbf{x^{*}}$
from the received solution $\mathbf{y}^{*}$, the end-user has to
calculate $\mathbf{Ky^{*}}$. Among those operations, the matrix multiplication
$\mathbf{AK}$ is the most computationally expensive one. Thus in
our discussion, we will analyze the \textcolor{black}{number of multiplications
$M$} required to compute $\mathbf{AK}$\textcolor{blue}{.} In the
following analysis, we denote $\mathbf{A}=\{a_{ij}\vert i,j=1,2,\cdots,n\}$
and $\mathbf{K}=\{k_{ij}\vert i,j=1,2,\cdots,n\}$. 

To multiply two arbitrary $n\times n$ matrices, the typical complexity
is $\mathcal{O}(n^{3})$, which is generally believed to be too high
and unacceptable for mobile client computation. However, in our design,
we can actually control the complexity by selecting matrix $\mathbf{K}$
properly so that the computational complexity can be effectively reduced
without compromising security. Since matrix multiplication is the
most expensive part of the end-user's processing, our goal is to ensure
that the complexity of multiplying $\mathbf{K}$ with an arbitrary
matrix $\mathbf{A}$ is bounded by $\mathcal{O}(n^{2})$, which is
within the end-user's computational constraints. 

In the following sections, we provide four schemes with different
types of non-singular secret key $\mathbf{K}$ based on the above
described complexity constraints.

\subsubsection{$\mathbf{K}$ is a Diagonal Matrix (Scheme-1)}

A diagonal matrix $\mathbf{K}$ has the format $\mathbf{K}=\{k_{ij}\vert k_{ij}=0,\forall i\neq j\}$.
 Since $\mathbf{K}$ must be non-singular, all the entries in the
diagonal have to be non-zero numbers.  When $\mathbf{K}$ is a diagonal
matrix, we have $M=n^{2}$.

\subsubsection{$\mathbf{K}$ is a Permutation Matrix (Scheme-2)}

A permutation matrix $\mathbf{K}$ has exactly one non-zero entry
in each row and each column in the matrix.  When $\mathbf{K}$ is
a permutation matrix, we have $M=n^{2}$.

\subsubsection{$\mathbf{K}$ is a Band Matrix (Scheme-3)}

Suppose the band matrix $\mathbf{K}$ has an upper half-bandwidth
$p$ and a lower half-bandwidth $q$ such that $k_{ij}=0$ for $i>j+p$
and $j>i+q$.  The total bandwidth of $\mathbf{K}$ is denoted by
$W=p+q+1$.  When $\mathbf{K}$ is a band matrix, for simplicity,
we assume that $\mathbf{K}$ has an equal upper and lower half-bandwidth
$p=q=\omega$, then $W=2\omega+1$, and the number of multiplications
$M$ can be calculated as $M=(2\omega+1)n^{2}-(\omega^{2}+\omega)n$.

\subsubsection{$\mathbf{K}$ is a sparse matrix (Scheme-4)}

Suppose $\mathbf{K}$ is a sparse matrix. The density $d$ is defined
as the ratio of non-zero elements in the matrix. We assume that the
number of non-zero elements in each row and each column of $\mathbf{K}$
is up-bounded by a constant $\theta$. When $\mathbf{K}$ is a sparse
matrix, it is usually stored in a special manner such as Dictionary
of Keys (DOK) \cite{pissanetzky1984sparse} in computation. Thus the
complexity of matrix multiplication can be approximately measured
by the number of non-zero elements, which is $dn^{3}$ in our discussion.
Since we have assumed that $d\leq\frac{\theta}{n}$, the number of
multiplication becomes $M=\theta n^{2}$. 

In summary, through the above analysis, we demonstrate that for the
four proposed schemes, the complexity of multiplying $\mathbf{K}$
with an arbitrary matrix $\mathbf{A}$ is $\mathcal{O}(n^{2})$. Since
matrix multiplication is the most expensive part of the end-user's
processing, we can derive that the overall computational complexity
for the end-user is $\mathcal{O}(n^{2})$, which is within the end-user's
computational constraints. 

For the four types of matrices, sparse matrix is the most general
case. When the non-zero elements are centralized around the diagonal,
then the sparse matrix $\mathbf{K}$ becomes a band matrix. When $\theta=W$,
the complexity of scheme-3 and scheme-4 is in the same level. The
only difference is that the non-zero elements are randomly distributed
in sparse matrix. This difference may provide different security protection
for side information which we will explain in detail later. Further,
when $\theta=1$, then $\mathbf{K}$ becomes a permutation matrix
as in scheme-2. Similarly when $W=1$, then the band matrix $\mathbf{K}$
becomes a diagonal matrix. Generally speaking, from scheme-1 to scheme-4,
the computational complexity increases. In the following sections,
we will analyze the security of CASO.

\subsection{Security Analysis\label{Sec:Security}}

In this section, we will analyze the security of our proposed CASO.
We will focus on the security of the coefficient matrix $\mathbf{A}$
of the original function $F(\mathbf{x})$, the variable $\mathbf{x}$
in the function $F(\mathbf{x})$ and the form of the function $F(\mathbf{x})$. 
\begin{thm}
For the four schemes in CASO, it is computationally infeasible for
the cloud to recover the original coefficient matrix $A$ of problem
$F(\mathbf{x})$ and the output $\mathbf{x}^{\ast}$ for the system
of linear equations. \end{thm}
\begin{IEEEproof}
For a system of linear equations $\mathbf{Ax}=\mathbf{b}$, the original
problem is represented by the matrix $\mathbf{A}$ and the vector
$\mathbf{b}$. The output is $\mathbf{x^{*}}$, which is the solution
of the system. Under the affine mapping, the system of equations is
transformed to $\mathbf{A^{\prime}}\mathbf{y}=\mathbf{b^{\prime}}$,
where $\mathbf{A^{\prime}}=\mathbf{AK}$ and $\mathbf{b^{\prime}}=\mathbf{b}-\mathbf{Ar}$.
Therefore, both $\mathbf{A}$ and $\mathbf{b}$ are concealed by the
secret key $\mathbf{S}=\{\mathbf{K},\mathbf{r}\}$. Since both $\mathbf{K}$
and $\mathbf{r}$ are only used once and kept secret at the local
side, the equations can be concealed form the cloud. Additionally,
since the original solution is recovered by $\mathbf{x}^{*}=\mathbf{K}\mathbf{y}^{*}+\mathbf{r}$,
without knowing $\mathbf{K}$ and $\mathbf{r}$, the cloud cannot
recover $\mathbf{x}^{*}$. In this way, the output of the system is
concealed. Thus, all the four schemes are secure in outsourcing the
system of linear equations. \end{IEEEproof}
\begin{thm}
\label{thm:indirect-var}Under the proposed CASO scheme, it is computationally
infeasible for the cloud to recover the zeros, poles and optimums. \end{thm}
\begin{IEEEproof}
Under the affine mapping $\mathbf{x}=\mathbf{Ky}+\mathbf{r}$, the
values of $\mathbf{x}$ is being mapped to $\mathbf{y}=\mathbf{K}^{-1}(\mathbf{x}-\mathbf{r}).$
Since both $\mathbf{r}$ and $\mathbf{K}$ are secret and randomly
selected, it is computationally infeasible to recover $\mathbf{x}$
from $\mathbf{y}$. As a result, given $\mathbf{y}$, the zeros, poles
and optimums are concealed. The protection depends on the selection
of $\mathbf{K}$ and $\mathbf{r}$. 
\end{IEEEproof}
While we cannot recover the coefficient matrix $A,$ the four scheme
do provide different levels of security protection for side information.
For scheme-1, the zeros of the coefficient matrix $A'$ of the outsourced
problem are also the zeros of the original problem. The order of the
entries in each column of $A'$ is the same as that of $A.$ For scheme-2,
while the number of zeros for $A'$ and $A$ are the same in each
column, the distribution of the zeros and the order of the entries
in each column are different. For scheme-3 and scheme-4, both the
number of zeros and the distribution of entries in each column of
$A'$ and $A$ are different. Therefore, the end user should select
the scheme based on whether the possible side information that may
be leaked from scheme-1 and scheme-2 is sensitive. 
\begin{thm}
\label{thm:indirect-form}Suppose $\psi$ is a rational mapping, meaning
that $\psi$ can be represented as a quotient of two polynomial functions,
$G=F\circ\psi$, then we have the following results:
\begin{enumerate}
\item If $F$ is a rational function, then $G$ is rational. 
\item If $F$ is an irrational function, then $G$ is irrational. 
\end{enumerate}
\end{thm}
\begin{IEEEproof}
Since $\psi$ is a rational mapping, we assume $\psi(x)=\frac{P(x)}{Q(x)}$,
where $P(x)$ and $Q(x)$ are polynomials. When $F$ is a rational
function, suppose 
\[
F(x)=\frac{a_{0}+a_{1}x+\cdots+a_{n}x^{n}}{b_{0}+b_{1}x+\cdots+b_{m}x^{m}}.
\]
Then
\[
{\displaystyle (F\circ\psi)(x)={\displaystyle \frac{a_{0}+a_{1}\frac{P(x)}{Q(x)}+\cdots+a_{n}\frac{P^{n}(x)}{Q^{n}(x)}}{b_{0}+b_{1}\frac{P(x)}{Q(x)}+\cdots+b_{m}\frac{P^{m}(x)}{Q^{m}(x)}}}.}
\]
Without loss of generality, we assume that $m>n$. Then we have 
\[
(F\circ\psi)(x)=\frac{a_{0}Q^{m}(x)+a_{1}P(x)Q^{m-1}(x)+\cdots+a_{n}P^{n}(x)Q^{m-n}(x)}{b_{0}Q^{m}(x)+b_{1}P(x)Q^{m-1}(x)+\cdots+b_{m}P^{m}(x)},
\]
where $F\circ\psi$ is the quotient of two polynomials. Thus, the
composition $G=F\circ\psi$ is a rational function. 

When $F$ is irrational, the composition $G=F\circ\psi$ cannot be
rational. Otherwise, there exists an inverse rational function $\psi^{-1}$
such that $F=G\circ\psi^{-1}=F\circ\psi\circ\psi^{-1}$ becomes rational.
Hence, $G=F\circ\psi$ is irrational when $F$ is irrational. 
\end{IEEEproof}
Since the proposed affine mapping is rational, we have the following
corollary. 
\begin{cor}
\label{cor:partial}Under an affine mapping $\psi$, the rationality
of the function $G$ is the same as the original function $F$. 
\end{cor}
Theorem \ref{thm:indirect-form} and Corollary \ref{cor:partial}
state that the rationality of the function $F$ cannot be changed
through composition with a rational mapping or an affine mapping $\psi$.
That is, if the function $F$ is rational, after the composition $G=F\circ\psi$,
the transformed function $G$ is still rational. If $F$ is irrational,
$G$ is still irrational. As a result, the side information that is
related to the specific form of the function $F$ (e. g., $\sin(\cdot)$
or $\log(\cdot)$ ) may not be fully concealed by an affine mapping
or even a rational mapping. 

Now, we will analyze the side information that can be revealed by
the coefficient matrix $\mathbf{A}$ of the four schemes. Under an
affine mapping, the coefficient matrix $\mathbf{A}$ is transformed
to $\mathbf{AK}$. For scheme-1, the secret key $\mathbf{K}$ is a
diagonal matrix denoted by $\mathbf{K}_{1}=\{k_{ij}\vert k_{ij}=0,\forall i\neq j\}$.
The entry $a_{ij}^{\prime}$ in $\mathbf{A}^{\prime}$ can be calculated
as $a_{ij}^{\prime}=k_{ii}a_{ij}$. By investigating $\mathbf{A}^{\prime}$,
it is obvious that each column in $\mathbf{A}^{\prime}$ is related
in a simple way to that in $\mathbf{A}$ such that the $i^{th}$ column
in $\mathbf{A}^{\prime}$ is the multiplication of the $i^{th}$ column
in $\mathbf{A}$ with $k_{ii}$. In this way, only based on $\mathbf{A}^{\prime}$,
the cloud can easily know the ratio between any two entries within
the same column in $\mathbf{A}$. 

For $\mathbf{K}$ to be a permutation matrix in scheme-2, the difference
is that $\mathbf{A}^{\prime}$ in scheme-2 can be regarded as the
result of permuting the columns of $\mathbf{A}^{\prime}$ obtained
from scheme-1. Thus, although the cloud can get a knowledge of the
ratio between two entries in the same column of $\mathbf{A}$, it
is not sure which particular column those two entries belong to. Therefore,
scheme-2 is more secure than scheme-1 in terms of concealing the side
information of the outsourced problem. 

In scheme-3, for $\mathbf{K}$ to be a band matrix with upper half-bandwidth
and lower half-bandwidth both equal to $\omega$, it can be calculated
that 
\[
a_{ij}^{\prime}=\sum_{r=j-\omega}^{j+\omega}a_{ir}k_{rj}.
\]
 Since each entry in $\mathbf{A}^{\prime}$ is a linear combination
of $\alpha_{3}$ entries in $\mathbf{A}$ and $\beta_{3}$ entries
in $\mathbf{K}$, the ratio information of entries in $\mathbf{A}$
is concealed. However, the disadvantage is that the cloud can still
learn how a particular entry in $\mathbf{A}^{\prime}$ is composed.
For example, suppose $\omega=1$, the cloud can know for sure that
$a_{ij}^{\prime}=a_{i(j-1)}k_{(j-1)j}+a_{ij}k_{jj}+a_{i(j+1)}k_{(j+1)j}$. 

At last, for $\mathbf{K}$ to be a sparse matrix in scheme-4, we assume
that there are exactly $\theta$ non-zero entries in each row and
column of $\mathbf{K}$. Similar to scheme-3, the ratio information
of entries in $\mathbf{A}$ can be concealed. Moreover, since the
non-zero entries are randomly positioned in the sparse matrix $\mathbf{K}$,
the cloud is unable to know how each entry in $\mathbf{A}^{\prime}$
is composed. 

From the above analysis, we can see that the coefficient matrix $\mathbf{A}$
is protected through the linear combination of the entries in $\mathbf{A}$
and $\mathbf{K}$. To be specific, in scheme-1 the effect is the scaling
of the columns of $\mathbf{A}$. In scheme-2, the effect is scaling
and permutation. In scheme-3 and scheme-4, the entries in $\mathbf{A}$
and $\mathbf{K}$ are related in a more complex way. We summarize
the computational complexity and security of CASO in Table \ref{Table:schemeSummary}. 

\begin{table}
\def\arraystretch{1.5}\centering \caption{Complexity and security of each scheme \label{Table:schemeSummary} }

\begin{tabular}{|c|c|}
\hline 
\textbf{Scheme } & \textbf{Complexity }\tabularnewline
\hline 
Diagonal matrix  & $n^{2}$ \tabularnewline
\hline 
Permutation matrix  & $n^{2}$ \tabularnewline
\hline 
Band matrix with bandwidth $W=(2\omega+1)$  & $Wn^{2}$ \tabularnewline
\hline 
Sparse matrix with density $d=\frac{\theta}{n}$  & $\theta n^{2}$ \tabularnewline
\hline 
\end{tabular}
\end{table}

\subsection{Trade-off between Complexity and Security}

From the above complexity and security analysis, we can see that there
is a trade-off between the computational complexity and security.
As the simple scheme, scheme-1 is able to protect the original coefficient
matrix while exposing the ratio between any two entries in the same
column. In comparison, scheme-2 is slightly more expensive (e. g.
the positions of the non-zero entries have to be stored), but it is
this cost for non-zero entries' random positions that makes it effective
to conceal the ratio information. The complexity of scheme-3 and scheme-4
is linearly dependent on $W$ and $\theta$, respectively. They are
more costly than scheme-1 and scheme-2. However, the transformed matrix
$\mathbf{A}^{\prime}$ can conceal $\mathbf{A}$ and $\mathbf{K}$
in a more complex way since it can conceal the structure of the coefficient
matrix. In summary, from scheme-1 to scheme-4, the security levels
that they can provide increase at a cost of computational power. 

In the context of cloud computing, the end-users vary from mobile
devices to powerful workstations thus having different computational
constraints as well as different security demands. Thus CASO provides
end-users with the flexibility to choose the outsourcing schemes that
are most suitable for them. These four schemes give cost-aware outsourcing
for end-users to address the various security demands and computational
constraints.

\subsection{Application to Linear Programming}

In this section, we will demonstrate that our design and analysis
for system of linear equations can be well applied to many computational
problems, such as linear programming.  We consider a linear programming
problem denoted by
\[
F(\mathbf{x}):=\begin{cases}
\text{minimize} & \mathbf{c}^{T}\mathbf{x}\\
\text{subject to} & \mathbf{Ax}=\mathbf{b}\\
 & \mathbf{Dx}\geq\mathbf{0},
\end{cases}
\]
where $\mathbf{b},\mathbf{c}\in\mathbb{R}^{n}$, $\mathbf{A}\in\mathbb{R}^{m\times n}$
and $\mathbf{D}\in\mathbb{R}^{s\times n}$ ($m,s\leq n$). 

Under the affine mapping $\mathbf{x}=\mathbf{Ky}+\mathbf{r}$, the
problem is transformed to 
\[
G(\mathbf{y}):=\begin{cases}
\text{minimize} & \mathbf{c}^{T}\mathbf{Ky}+\mathbf{c}^{T}\mathbf{r}\\
\text{subject to} & \mathbf{AKy}=\mathbf{b-Ar}\\
 & \mathbf{DKy}\geq\mathbf{-Dr},
\end{cases}
\]
from which we can see that the original coefficient matrix can be
concealed by the secret key $\mathbf{K}$ and $\mathbf{r}$. It is
obvious that the computational bottleneck lies in the multiplication
of $\mathbf{K}$ with $\mathbf{A}$ and $\mathbf{D}$. Thus the same
complexity and security analysis for systems of linear equations applies
for linear programming. That is the complexity of the previous four
schemes is all bounded by $\mathcal{O}(n^{2})$. In terms of security,
the four schemes are all secure in protecting the original coefficient
matrix while providing different levels of protection of the side
information. 

In the next section, we explore the differences for non-linear computation
by investigating system of non-linear equations and convex optimization
problems.

\section{Extension to Non-linear Systems\label{sec:Extension-to-Non-linear}}

In this section, we aim at exploring the different design issues between
linear and non-linear computation. We consider a system of non-linear
equations denoted by $F(\mathbf{x})=\mathbf{0}$, where $F(\mathbf{x})=\{f_{i}(\mathbf{x})\vert f_{i}(\mathbf{x}):\mathbb{R}^{n}\rightarrow\mathbb{R},i=1,2,\cdots,n\}$.
Typically, it is hard to obtain a symbolic solution for the system.
Thus the normal method is to solve the system of equations numerically
in an iterative way. The main idea is that given a solution $\mathbf{x}_{k}$
in the $k^{th}$ iteration, we need to solve the linear system $\partial F(\mathbf{x})\vert_{\mathbf{x}=\mathbf{x}_{k}}(\mathbf{x}_{k+1}-\mathbf{x}_{k})=-F(\mathbf{x})\vert_{\mathbf{x}=\mathbf{x}_{k}}$,
where $\partial F(\mathbf{x})$ is the Jacob matrix of $F(\mathbf{x})$.
Then we can obtain the solution $\mathbf{x}_{k+1}$ in the $(k+1)^{th}$
iteration. The iteration will terminate when $\|F(\mathbf{x}^{*})\|<\varepsilon$,
where $\varepsilon$ is the \textcolor{black}{error tolerance} and
$\mathbf{x}^{*}$ is the final solution. To minimize the communication
overhead and the energy consumption of the end-users, our goal is
to design off-line scheme so that the end-users are not required to
interact with the cloud except the problem outsourcing and result
retrieving process. In this way, the end-users only need to focus
on the high level view of the problem without knowing the details
of problem solving process. The detailed design and analysis of the
outsourcing scheme are presented as follows.

\subsection{Outsourcing Scheme\label{sub:Outsourcing-Scheme}}

Compared with outsourcing of the system of linear equations, the main
difference lies in the problem transformation phase. First, to start
the iteration at the cloud side, an initial guess of the solution
should also be outsourced. We assume that at the local side, the end-user
generates an initial solution $\mathbf{x}_{0}$. Then with the affine
mapping, the outsourced initial solution becomes $\mathbf{y}_{0}=\mathbf{K}^{-1}(\mathbf{x}_{0}-\mathbf{r})$.
We should notice that there is an inversion operation on $\mathbf{K}$
which will impose more constraints on our selection of $\mathbf{K}$
in terms of computational complexity. Second, after substituting $\mathbf{x}$
with $\mathbf{y}$, the problem should be further transformed. We
use a simple example to illustrate this point. Suppose we want to
solve a system of nonlinear equations

\begin{equation}
F(\mathbf{x}):=\left\{ \begin{array}{l}
\sin(3x_{1})+4x_{2}^{2}+x_{2}x_{3}=0\\
2x_{1}+e^{3x_{2}}+2x_{3}^{3}=0\\
\lg(5x_{1})+\frac{1}{2x_{2}+1}+3(x_{3}+1)^{2}=0.
\end{array}\right.
\end{equation}
We take the affine mapping $\mathbf{x}=\mathbf{K}\mathbf{y}+\mathbf{r}$,
where $\mathbf{r}=\mathbf{0}$ and
\[
\mathbf{K}=\begin{bmatrix}3 & 0 & 0\\
0 & 2 & 0\\
0 & 0 & 4
\end{bmatrix}.
\]
Then the system is transformed to 
\begin{equation}
G(\mathbf{y}):=\left\{ \begin{array}{l}
\sin(9y_{1})+16y_{2}^{2}+8y_{2}y_{3}=0\\
6y_{1}+e^{6y_{2}}+128y_{3}^{3}=0\\
\lg(15y_{1})+\frac{1}{4y_{2}+1}+48y_{3}+24y_{3}=-3.
\end{array}\right.
\end{equation}

It is obvious that to protect the cloud from revealing information
from the transformed system, it is sufficient to mix the coefficient
of each term in the equations with the key entry. To be specific,
we assume that there are $\pi_{i}$ terms in equation $f_{i}(\mathbf{x})$
and each term is denoted by $f_{i}^{j}(\mathbf{tx})$, where $\mathbf{t}$
is the coefficient. Then each equation in the system can be written
as
\[
f_{i}(\mathbf{x})=\sum_{j=1}^{\pi_{i}}f_{i}^{j}(\mathbf{tx}).
\]
Under the affine mapping $\mathbf{x}=\mathbf{Ky}+\mathbf{r}$, $f_{i}^{j}(\mathbf{tx})$
is transformed to
\[
g_{i}(\mathbf{y})=f_{i}(\mathbf{Ky+r})=\sum_{j=1}^{\pi_{i}}f_{i}^{j}(\mathbf{t(Ky+r)}).
\]
Thus the coefficient $\mathbf{t}$ is concealed by $\mathbf{K}$ and
$\mathbf{r}$, which is similar to the case of the system of linear
equations. However, as illustrated in the example, the multiplication
cannot be simply carried out when $f_{i}^{j}(\cdot)$ is a polynomial.
Thus a further transformation is needed to mix $\mathbf{t}$ with
$\mathbf{K}$ and $\mathbf{r}$ for polynomials. 

Without loss of generality, we assume that the polynomial is denoted
by $t_{i}x_{i}^{m}$ and in the affine mapping, $\mathbf{K}$ is a
band matrix with bandwidth $W=3$ and $\mathbf{r}=\mathbf{0}$. Thus
under the affine mapping, the polynomial is transformed to $t_{i}(k_{i-1}y_{i-1}+k_{i}y_{i}+k_{i+1}y_{i+1})^{m}$.
To mix the coefficient $t_{i}$ with the secret keys, one straightforward
way is to expand the polynomial and then multiple it with $t_{i}$.
However, the complexity is unacceptable for high order polynomials.
Instead, we propose that it is sufficient to split the secret keys
as $k_{s}=pq_{s}$, where $s=i-1,i,i+1$ such that $t_{i}(k_{i-1}y_{i-1}+k_{i}y_{i}+k_{i+1}y_{i+1})^{m}=t_{i}(pq_{i-1}y_{i-1}+pq_{i}y_{i}+pq_{i+1}y_{i+1})^{m}=t_{i}p^{m}(q_{i-1}y_{i-1}+q_{i}y_{i}+q_{i+1}y_{i+1})^{m}$.
In this way, the coefficient $t_{i}$ in the original function and
the secret keys $k_{i}$ are concealed.

\subsection{Complexity Analysis}

From the analysis above, we can see that the complexity of the problem
transformation mainly depends on two aspects. One is the specific
form of the equations, that is the number of polynomials in the equations.
The other one is how $\mathbf{x}$ and $\mathbf{y}$ are related,
which is determined by the number of non-zero entries in $\mathbf{K}$. 

For a given system of non-linear equations, suppose that there are
$N$ terms in total in the systems, among which $L$ are polynomials
with orders no greater than $m$. Assume that the number of non-zero
entries in $\mathbf{K}$ is up-bounded by $\lambda$ (i. e. each $x$
is substituted by at most $\lambda$ $y$'s). Thus for each non-polynomial
term, the transformation takes $\lambda$ multiplications between
the coefficient of the term and the key entries. And for a polynomial
term $t_{i}x_{i}^{m}$, we assume that it is replaced by $t_{i}(k_{1}y_{1}+\cdots+k_{\lambda}y_{\lambda})^{m}=t_{i}(pq_{1}y_{1}+\cdots+pq_{\lambda}y_{\lambda})^{m}=t_{i}p^{m}(q_{1}y_{1}+\cdots+q_{\lambda}y_{\lambda})^{m}$.
Then the operations involved in the transformation include one multiplication,
$\lambda$ division and raising $p$ to the power of $m$. As stated
previously, we utilize the number of multiplication as a measurement
for complexity. We assume that in terms of computational complexity,
one division is equal to one multiplication and with the method of
exponentiation by squaring, the computation for $m^{th}$ power takes
$\log^{2}m$ multiplications. Thus, for a system of non-liner equations
with $N$ terms among which $L$ are polynomials, the complexity can
be calculated as 
\[
\lambda N+(\log^{2}m+1)L.
\]
It is obvious that the complexity depends on $\lambda$ which is further
determined by the selection of $\mathbf{K}$. We summarize the complexity
of the four different types of matrices in Table \ref{Table:Non-linear Complexity}.
We can see from the table that the complexities of all schemes are
constrained to $\mathcal{O}(N)$, where $N$ is the number of terms
in the system of non-linear equations. Notice that typically for a
system of equations, the number of terms $N$ is in the level of $n^{2}$,
where $n$ is the number of independent variables. Thus the complexity
is still bounded by $\mathcal{O}(n^{2})$, which fulfills our design
goals. 

\begin{table}
\def\arraystretch{1.5}\centering \caption{Complexity for system of non-linear equations \label{Table:Non-linear Complexity} }

\begin{tabular}{|c|c|}
\hline 
\textbf{Scheme } & \textbf{Complexity }\tabularnewline
\hline 
Diagonal matrix  & $N+(\log^{2}m+1)L$ \tabularnewline
\hline 
Permutation matrix  & $N+(\log^{2}m+1)L$ \tabularnewline
\hline 
Band matrix with bandwidth $W=(2\omega+1)$  & $WN+(\log^{2}m+1)L$ \tabularnewline
\hline 
Sparse matrix with density $d=\frac{\theta}{n}$  & $\theta N+(\log^{2}m+1)L$ \tabularnewline
\hline 
\end{tabular}
\end{table}

\subsection{Security Analysis}

Similar to the security analysis for linear systems, all of the proposed
four schemes are secure in protecting the coefficient matrix, the
zeros, poles and optimums of the outsourced problem. As stated in
Corollary \ref{cor:partial}, CASO cannot conceal the specific form
of the functions. For instance, in the example given in Section \ref{sub:Outsourcing-Scheme},
the original system of equations is transformed to $G(\mathbf{y})$
such that the coefficients in each term of the function are changed.
However, the specific forms of the function (e. g., $\sin(\cdot)$,$\lg(\cdot)$,
etc. ) remain unchanged. 

For the four schemes, generally as the complexity increases, more
side information can be concealed from the cloud. Different from the
linear equations, a non-linear function $f_{i}(x)$ may contain some
side information, such as maximum or minimum value which is important
in some applications. For instance, the plot of the function or the
extreme values may expose the distribution of the incidence of a disease
among different age groups. For scheme-1 and scheme-2, the curve of
the function is just a scaled version. Though scheme-2 provides better
protection since it can conceal the independent variables. In scheme-3
and scheme-4, each independent variables in the original problem is
substituted by several new variables. Thus the side information, such
as the curve and the extreme values can be perfectly concealed.

\subsection{Application to Convex Optimization}

In this section, we show that the above schemes and analysis can also
be applied to convex optimization. Convex optimization is widely deployed
in various practical problems. We consider a convex optimization problem
denoted by
\begin{equation}
F(\mathbf{x}):=\begin{cases}
\text{minimize} & f_{0}(\mathbf{x})\\
\text{subject to} & f_{i}(\mathbf{x})\leq0,i=1,\cdots,m\\
 & h_{j}(\mathbf{x})=0,j=1,\cdots,t,
\end{cases}\label{eq:ConvexOpt}
\end{equation}
where $f_{i}:\mathbb{R}^{n}\rightarrow\mathbb{R}$, $i=0,...,m$ and
$h_{i}:\mathbb{R}^{n}\rightarrow\mathbb{R}$, $i=1,...,t$ are all
convex functions. Under the affine mapping $\mathbf{x}=\mathbf{Ky}+\mathbf{r}$,
the original problem $F(\mathbf{x})$ is transformed to
\[
G(\mathbf{y}):=\begin{cases}
\text{minimize} & f_{0}(\mathbf{Ky}+\mathbf{r})\\
\text{subject to} & f_{i}(\mathbf{Ky}+\mathbf{r})\leq0,i=1,\cdots,m\\
 & h_{j}(\mathbf{Ky}+\mathbf{r})=0,j=1,\cdots,t.
\end{cases}
\]

Since the key matrix $\mathbf{K}$ and $\mathbf{r}$ are randomly
generated and kept secret at the local side, the coefficient matrix
of the outsourced problem is perfectly protected. And because the
functions $f_{i}(\cdot)$ and $h_{j}(\cdot)$ are all non-linear functions,
the security and the complexity analysis of system of non-linear equations
can be well applied in this case. Thus we conclude that our outsourcing
scheme is also applicable to convex optimization problems.

\section{Results Verification\label{Sec:Results-Verification}}

In this section, we propose a result verification scheme under which
the end-users are guaranteed to receive the valid results from the
cloud. As stated previously, the cloud is not only curious about the
end-users' private information, it may also behave ``lazily'' to
increase its own benefits. That is, to reduce the computational cost,
the cloud has the motivation to deploy less computational resources
and simply returns some trivial results. As a consequence, the end-users
are not able to recover the valid results from those returned by the
cloud. 

The general idea of our proposed verification scheme is to outsource
the problem twice under two different affine mappings and to verify
whether the two results returned by the cloud match with each other.
To be specific, under the affine mappings $\mathbf{x}=\mathbf{K}_{1}\mathbf{y}+\mathbf{r}_{1}$
and $\mathbf{x}=\mathbf{K}_{2}\mathbf{z}+\mathbf{r}_{2}$, the original
problem $F(\mathbf{x})$ is transformed to $G(\mathbf{y})$ and $H(\mathbf{z})$
which are outsourced to the cloud. Then the cloud solves the two outsourced
problems and returns the corresponding results $\mathbf{y}^{*}$ and
$\mathbf{z}^{*}$. Since the condition $\mathbf{K}_{1}\mathbf{y}^{*}+\mathbf{r}_{1}=\mathbf{K}_{2}\mathbf{z}^{*}+\mathbf{r}_{2}$
holds for these two results, the end-users can utilize it as a criterion
to verify whether the returned results are valid.

\subsection{System of Equations}

The idea introduced above can be applied to system of equations directly.
When $F(\mathbf{x})$ is a system of linear equations, it is sufficient
to verify directly whether $\|\mathbf{A}\mathbf{x}^{*}\|<\varepsilon$,
where $\|\cdot\|$ denotes the Euclidean norm of a vector and $\varepsilon$
is a pre-defined\textcolor{black}{{} error tolerance}. The complexity
of this verification process is $\mathcal{O}(n^{2})$. 

When $F(\mathbf{x})$ is a system of non-linear equations, since the
end-user will have to evaluate the non-linear functions, the computational
cost for direct verification generally exceeds $\mathcal{O}(n^{2})$.
However, based on our idea of outsourcing twice, the end-user only
needs to check the condition $\mathbf{K}_{1}\mathbf{y}^{*}+\mathbf{r}_{1}=\mathbf{K}_{2}\mathbf{z}^{*}+\mathbf{r}_{2}$.
Since the verification process involves only linear operations, the
computational complexity is bounded by $\mathcal{O}(n^{2})$. As system
of equations is typically solved by iterative method, the solution
is not accurate. Thus we may need to change the equality condition
to $\|\mathbf{(K}_{1}\mathbf{y}^{*}+\mathbf{r}_{1})-(\mathbf{K}\mathbf{z}^{*}+\mathbf{r}_{2})\|<\varepsilon$.
In the following analysis, we uniformly utilize the equality condition
$\mathbf{K}_{1}\mathbf{y}^{*}+\mathbf{r}_{1}=\mathbf{K}_{2}\mathbf{z}^{*}+\mathbf{r}_{2}$
as the verification criteria. When the computational problems are
solved inaccurately, the equality condition should be changed to its
inequality variation.

\subsection{Optimization Problems}

When $F(\mathbf{x})$ is an optimization problem, we utilize convex
optimization as an example to illustrate the verification process.
And it can be easily applied to other optimization problems, such
as linear programming. The output of a convex optimization problem
can be divided into three cases: normal, infeasible and unbounded
\cite[Chapter 4.1]{boyd2009convex}. For the convex optimization problem
defined in equation (\ref{eq:ConvexOpt}), the \emph{domain} $\mathbb{\mathcal{D}}$
is the set for which the objective function and the constraint functions
are defined. That is
\[
\mathcal{D}=\bigcap_{i=1}^{m}\mbox{\textbf{dom}}f_{i}\cap\bigcap_{i=i}^{t}\mbox{\textbf{dom}}h_{i}.
\]
The \emph{feasible set} is $\mathcal{E}=\left\{ \mathbf{x}\in\mathcal{D\mid}f_{i}(\mathbf{x})\leq0,i=1,\cdots,m,h_{i}(\mathbf{x})=0,i=1,\cdots,t\right\} $.
In the normal case, there exists an optimal point $\mathbf{x}^{*}\in\mathcal{E}$
such that $f_{0}(\mathbf{x^{\mathbf{*}}})\leq f_{0}(\mathbf{x)},\forall\mathbf{x}\in\mathcal{E}$.
In the infeasible case, $\mathcal{E=\textrm{\ensuremath{\emptyset}}}$.
In the unbounded case, there exists points $\mathbf{x}_{k}\in\mathcal{E}$
such that $f_{0}(\mathbf{x}_{k})\rightarrow-\infty$ as $k\rightarrow\infty$. 

For the cloud to cheat, it must return results in the same case for
the two outsourced problem $G(\mathbf{y})$ and $H(\mathbf{z})$ as
mentioned above.  Suppose that $\mathbf{y}^{*}$ and $\mathbf{z}^{*}$
are the two returned results and they belong to the same case.  In
the following, we will present the verification scheme for the three
different cases separately.

\subsubsection{Normal Case}

The above proposed verification scheme works well for the normal case.
 That is if the equality $\mathbf{K}_{1}\mathbf{y}^{*}+\mathbf{r}_{1}=\mathbf{K}_{2}\mathbf{z}^{*}+\mathbf{r}_{2}$
holds, the end-user can make sure that a valid result can be recovered.
 This is because whatever the correct result is (normal, infeasible
or unbounded), the cloud is not able to come up with two results that
satisfy the equality without actually conduct the computation process.
 And this verification process for normal case forms the basis for
the verification for other cases.

\subsubsection{Infeasible Case }

The above verification scheme would fail if the cloud simply returns
an infeasible result for any outsourced convex optimization problem.
 To deal with this issue, we utilize phase I method as described in
\cite[Chapter 11]{boyd2009convex} to check the feasibility of the
problem.  For a convex optimization problem $F(\mathbf{x})$, a corresponding
phase I optimization problem can be constructed as:
\[
F_{I}(\mathbf{x}):=\left\{ \begin{array}{ll}
\text{minimize} & \rho\\
\text{subject to} & f_{i}(\mathbf{x})\leq{\color{red}{\color{black}\rho}},i=1,\cdots,m\\
 & h_{j}(\mathbf{x})=0,j=1,\cdots,t
\end{array}\right.,
\]
where $\rho$ is a single variable.  It is obvious that when $\rho$
is large enough, $F_{I}(\mathbf{x})$ is always feasible.  

Suppose $\mathbf{x}^{*}$ minimizes the objective function and $\rho^{*}$
is the corresponding minimum value.  The phase I problem is designed
in such a way that when $\rho^{*}\leq0$, the original problem $F(\mathbf{x})$
is feasible and $F(\mathbf{x})$ is infeasible otherwise.  Thus the
verification scheme for infeasible case can be designed as follows.
 When the cloud indicates that the solutions to the two outsourced
problem $G(\mathbf{y})$ and $H(\mathbf{z})$ are infeasible, it then
generates the corresponding two phase I problems $G_{I}(\mathbf{y})$
and $H_{I}(\mathbf{z})$ and computes the optimal points $\mathbf{y}^{*}$
and $\mathbf{z}^{*}$ and the minimum values $\rho_{G}^{*}$ and $\rho_{H}^{*}$,
respectively.  Then at the local side, the verification is the same
as that in the normal case.  That is only when $\rho_{G}^{*}>0$ and
$\rho_{H}^{*}>0$ and the equality $\mathbf{K}_{1}\mathbf{y}^{*}+\mathbf{r}_{1}=\mathbf{K}_{2}\mathbf{z^{*}}+\mathbf{r}_{2}$
holds can the end-user be guaranteed to receive valid solutions.

\subsubsection{Unbounded Case}

In the unbounded case, the cloud indicates that the objective function
$f_{0}(\mathbf{x})\rightarrow-\infty$ in its domain. We utilize duality
to verify the soundness of the returned result. For a convex optimization
problem, we can construct the corresponding \emph{Lagrangian} $L$
as
\[
L(\mathbf{x},\mathbf{u},\mathbf{v})=f_{0}(\mathbf{x})+\sum_{i=1}^{m}u_{i}f_{i}(\mathbf{x})+\sum_{j=1}^{t}v_{j}h_{j}(\mathbf{x}),
\]
where $\mathbf{u}\in\mathbb{R}^{m}$ and $\mathbf{v}\mathbb{\in R}^{t}$
are the associated \emph{Lagrange multiplier vectors}. Then based
on this Lagrangian $L(\mathbf{x},\mathbf{u},\mathbf{v})$, a \emph{Lagrange
dual function} can be constructed as 
\begin{eqnarray*}
\Phi(\mathbf{u},\mathbf{v}) & = & \inf_{\mathbf{\mathbf{x}\in\mathcal{D}}}L(\mathbf{x},\mathbf{u},\mathbf{v})\\
 & = & \underset{\mathbf{x}\in\mathcal{D}}{\inf}\left(f_{0}(\mathbf{x})+\sum_{i=1}^{m}u_{i}f_{i}(\mathbf{x})+\sum_{j=1}^{t}v_{j}h_{j}(\mathbf{x})\right),
\end{eqnarray*}
where $\mathcal{D}$ is the domain of the optimization problem. From
this definition, it is easy to prove that $\forall\mathbf{u}\succeq0$,
we have the following inequality:
\[
\Phi(\mathbf{u},\mathbf{v})\leq L(\mathbf{x}^{*},\mathbf{u},\mathbf{v})\leq f_{0}(\mathbf{x}^{*}),
\]
where $f_{0}(\mathbf{x}^{*})$ denotes the optimal value of the objective
function. The above inequality gives a lower bounded of the objective
function that depends on the selection of $\mathbf{u}$ and $\mathbf{v}$.
Thus, among all the selections of $\mathbf{u}$ and $\mathbf{v}$,
find the optimal lower bound is equivalent to solving the following
optimization problem:
\[
\begin{cases}
\text{maximize} & \Phi(\mathbf{u},\mathbf{v})\\
\text{subject to} & \mathbf{u}\succeq0.
\end{cases}
\]
The objective function $\Phi(\mathbf{u},\mathbf{v})$ is concave since
it is the point-wise infimum of a series of affine function of $(\mathbf{u},\mathbf{v})$.
Thus the above optimization problem is also a convex optimization
problem. If the original problem is unbounded below, the convex optimization
problem described above should be infeasible since it gives a lower
bound of the optimal value in the original problem. Thus the remaining
task is to verify the feasibility of the above convex optimization
problem, which has been illustrated in the infeasible case. Let the
cloud solve the phase I problems of the two Lagrange dual problems
and return the optimal solutions denoted by $(\rho_{G}^{*},\mathbf{y}^{*},\mathbf{u}_{G}^{*},\mathbf{v}_{G}^{*})$
and $(\rho_{H}^{*},\mathbf{z}^{*},\mathbf{u}_{H}^{*},\mathbf{v}_{H}^{*})$.
At the local side, the end-user then checks whether $\rho_{G}^{*}>0$
and $\rho_{H}^{*}>0$ and whether the equality $\mathbf{K}_{1}\mathbf{y}^{*}+\mathbf{r}_{1}=\mathbf{K}_{2}\mathbf{z}^{*}+\mathbf{r}_{2}$
holds.

\section{Evaluation\label{sec:Evaluation}}

In this section, we will evaluate the performance of the proposed
CASO scheme. We first compare CASO with several existing outsourcing
schemes. Then we present some numeric results to show the efficiency
of CASO.

\subsection{Performance Comparison}

The existing schemes on outsourcing of numeric computation mainly
focus on some specific problems. To the best of our knowledge, no
effective outsourcing schemes have been proposed for non-linear problems.
For instance, in \cite{wang2011infocom} and \cite{wang2011icdcs},
the authors proposed two outsourcing schemes specially designed for
linear programming and system of linear equations, respectively. In
\cite{xu2013infocommini}, the authors focus on the result verification
of convex optimization problems without giving an outsourcing scheme.
In comparison, we propose an outsourcing scheme that is suitable for
general computational problems, including the problems investigated
in the above works. Especially, we present the application of our
scheme on both linear and non-linear problems. 

In the following part, we compare the performance of our proposed
CASO scheme with three existing schemes specially designed for three
types of problems in terms of security, computational complexity and
communication overhead. To measure the communication overhead, we
introduce a \emph{communication overhead index }$\mathcal{I}_{c}$\emph{
}which is defined as the fraction of the communication cost of transmitting
the original problem over that of the transformed problem. Thus a
larger $\mathcal{I}_{c}$ indicates better communication efficiency.

\subsubsection{Linear Programming}

In this section, we compare CASO for linear programming problems with
the schemes proposed in \cite{wang2011infocom} and \cite{znie2014efficient}
in both security and complexity. We will show that while achieving
the same security level, our scheme outperforms in terms of complexity.
In addition, our scheme also provides end-users with the flexibility
to select different outsourcing options with different complexity
according to their security demands. 

The general linear programming problems can be expressed as 
\begin{equation}
\begin{cases}
\text{minimize} & \mathbf{c}^{\mathbf{T}}\mathbf{x}\\
\mbox{subject to} & \mathbf{Ax}=\mathbf{b}\\
 & \mathbf{Dx}\geq\mathbf{0}.
\end{cases}\label{eq:Linear-Programming}
\end{equation}

In \cite{wang2011infocom}, to transform the problem, a secret key
$\mathbf{K}=\{\mathbf{Q},\mathbf{\mathbf{M},r,\boldsymbol{\lambda}},\gamma\}$
is generated, where $\mathbf{Q}$ is a randomly generated $m\times m$
non-singular matrix, $\mathbf{M}$ is a randomly generated $n\times n$
non-singular matrix, and $\mathbf{r}$ is an $n\times1$ vector. With
this secret key, the original problem is transformed to the following
problem
\[
\begin{cases}
\text{minimize} & \mathbf{c}'^{\mathbf{T}}\mathbf{x}\\
\mbox{subject to} & \mathbf{A'x}=\mathbf{b'}\\
 & \mathbf{D'x}\geq\mathbf{0},
\end{cases}
\]
where $\mathbf{A}^{\prime}=\mathbf{QAM},\mathbf{D}^{\prime}=(\mathbf{D}-\mathbf{\lambda}\mathbf{QA})\mathbf{M},\mathbf{b}^{\prime}=\mathbf{Q}(\mathbf{b}+\mathbf{Ar})$
and $\mathbf{c}^{\prime}=\gamma\mathbf{M}^{\mathbf{T}}\mathbf{c}$.
Then the transformed problem is outsourced to the cloud which is similar
as our approach. 

In terms of computational complexity, the computational overhead of
the outsourcing scheme in \cite{wang2011infocom} as well as our scheme
lies primarily in matrix multiplication. As stated in their paper,
the overall computational complexity for the scheme proposed in \cite{wang2011infocom}
is slightly less than $\mathcal{O}(n^{3})$ depending the algorithm
chosen to implement matrix multiplication. For instance, when the
Strassen algorithm is adopted, the complexity becomes $\mathcal{O}(n^{2.81})$;
while for the Coppersmith-Winograd algorithm the complexity is $\mathcal{O}(n^{2.376})$.
However, by carefully selecting the secret key $\mathbf{K}$, our
scheme can limit the complexity within $\mathcal{O}(n^{2})$. 

In terms of communication overhead, the original problems in both
schemes are transformed by matrix multiplication such that the resulting
matrices are still in the same scale. As a result, the communication
cost of the original and transformed problems are in the same level.
Thus we have $\mathcal{I}_{c}=1$ in our scheme and the scheme in
\cite{wang2011infocom}. 

In terms of security, both schemes can conceal the private information
by some disguising techniques, that is to disguise the original matrices
by multiplying them with some random matrices. As a consequence, the
security they can achieve in protecting the original coefficient matrix
is in the same level. Since the types of the transformation matrices
(e. g. $\mathbf{Q}$, $\mathbf{M}$) are not specified, each entry
in the disguised coefficient matrix $\mathbf{A}^{\prime}$ can be
the linear combination of multiple entries in $\mathbf{A}$ and the
transformation matrices. Thus, the ratio information can be concealed.
In this sense, the security of the scheme in \cite{wang2011infocom}
is comparable with our scheme-4 in terms of protecting side information. 

The scheme proposed in \cite{znie2014efficient} can be regarded as
a variation of that in \cite{wang2011infocom}. The main difference
is that the authors in \cite{znie2014efficient} specify the transformation
matrices as sparse matrices in to order to achieve a lower computational
complexity of $\mathcal{O}(n^{2})$. For example, the schemes in \cite{znie2014efficient}
disguises the coefficient matrix by matrix multiplication as $\mathbf{A}^{\prime}=\mathbf{MAN}$,
where $\mathbf{M}$ and $\mathbf{N}$ are both sparse matrices. In
this way, the complexity is reduced to $\mathcal{O}(n^{2})$. Actually,
this scheme can be considered as a special case of our proposed CASO
where $\mathbf{K}$ is selected as a sparse matrix.

\subsubsection{System of Linear Equations \label{sub:System-of-Linear}}

In \cite{wang2011icdcs}, the authors investigated outsourcing of
system of linear equations $\mathbf{Ax}=\mathbf{b}$ based on iterative
method. First, the problem is transformed to $\mathbf{Ay}=\mathbf{b^{\prime}}$,
where $\mathbf{y=x+r}$, $\mathbf{b^{\prime}=\mathbf{b}+\mathbf{Ar}}$
and $\mathbf{r}$ is a random vector. Then the end-user solves the
transformed problem iteratively with the aid of cloud servers and
an initial guess $\mathbf{y}_{0}$ from the following iteration equation:
\begin{equation}
\mathbf{y_{k+1}=\mathbf{T}\cdot\mathbf{y_{k}+c^{\prime},}}
\end{equation}
where $\mathbf{A=D+R}$ such that $\mathbf{D}$ is non-singular, $\mathbf{T}=-\mathbf{D^{-1}\cdot\mathbf{R}}$
and $\mathbf{c^{\prime}=\mathbf{D^{-1}\cdot\mathbf{b^{\prime}}}}$.
The end-user utilizes the cloud servers to compute the most expensive
part $\mathbf{T}\cdot\mathbf{y_{k}}$ based on homomorphic encryption
to conceal the private information $\mathbf{T}$. To be specific,
the matrix $\mathbf{T}$ is pre-computed at the local side and the
encrypted version $\mathsf{Enc}(\mathbf{T})$ is outsourced to the
cloud. At each iteration, the end-user sends $\mathbf{y_{k}}$ to
the cloud and based on the homomorphic properties of the encryption,
the cloud servers compute $\mathsf{Enc}(\mathbf{T\cdot y_{k}})$ by
\[
\begin{array}{ccl}
\mathsf{Enc}(\mathbf{T\cdot y_{k}})[i] & = & \mathsf{Enc}(\sum_{j=1}^{n}\mathbf{T}[i,j]\cdot y_{k,j})\\
 & = & \prod_{j=1}^{n}\mathsf{Enc}(\mathbf{T}[i,j])^{y_{k,j}}
\end{array}
\]
for $i=1,\cdots,n$ and send $\mathsf{Enc}(\mathbf{T\cdot y_{k}})$
back to the end-user. On receiving $\mathsf{Enc}(\mathbf{T\cdot y_{k}})$,
the end-user decrypts it and get $\mathbf{y_{k+1}}$. This iteration
terminates when it converges to the final result $\mathbf{y}$. At
last the end-user can recover the desired solution $\mathbf{x}$ by
$\mathbf{x=y}-\mathbf{r}$. 

As stated above, the computational overhead at the local side primarily
lies in the decryption of $\mathbf{T}\cdot\mathbf{y_{k}}$ in each
iteration. Suppose the algorithm terminates after $L$ rounds of iteration,
then the end-user has to perform $L\cdot n$ times of decryption.
However, the decryption process of public-key cryptosystem is much
more expensive than simple multiplication of real numbers since it
mainly consists of modular exponentiation of large numbers. For instance,
the decryption process \cite{paillier1999eurocrypt} adopted in \cite{wang2011icdcs}
has a complexity of $\mathcal{O}(n^{3})$ and a modified version can
achieve a complexity of $\mathcal{O}(n^{2+\epsilon})$. Thus, the
outsourcing scheme in \cite{wang2011icdcs} introduces $\mathcal{O}(n^{3+\epsilon})$
computational overhead at the local side. In terms of communication
overhead, the outsourcing process requires the end-user to send $\mathbf{y_{k}}$
and receive $\mathsf{Enc}(\mathbf{T\cdot y_{k}})$ at each iteration.
As a consequence, the communication overhead index $\mathcal{I}_{c}=\frac{1}{L}$
is dependent on the convergence speed. Furthermore, this iteration
process requires the end-user to be ``online'' for the process to
continue. In comparison, our scheme can limit the computational overhead
to $\mathcal{O}(n^{2})$ with $\mathcal{I}_{c}=1$. Moreover, during
the outsourcing process, the end-user is ``offline'', which means
that after outsourcing the transformed problem, the end-user does
not need to interact with the cloud servers until the result is sent
back. 

The system of linear equations considered in \cite{wang2011icdcs}
includes the coefficient matrix $\mathbf{T}$ and the solution vector
$\mathbf{x}$. In \cite{wang2011icdcs}, the matrix $\mathbf{T}$
is encrypted utilizing the Paillier cryptosystem\cite{paillier1999eurocrypt}
as $\mathsf{Enc}(\mathbf{T})$ and the vector $\mathbf{x}$ is transformed
to $\mathbf{y}=\mathbf{x}+\mathbf{r}$, where $\mathbf{r}$ is a random
vector. In comparison, CASO disguises the coefficient matrix $\mathbf{A}$
and the solution vector $\mathbf{x}$ as $\mathbf{A}^{\prime}=\mathbf{AK}$
and $\mathbf{x}=\mathbf{Ky}+\mathbf{r}$, respectively. In the Paillier
cryptosystem, each entry of the coefficient matrix $\mathbf{T}(i,j)$
is encrypted as $\mathsf{Enc}(\mathbf{T}(i,j))=g^{\mathbf{T}(i,j)}r^{n}\bmod n^{2}$,
where $g,r,n$ are parameters in the cryptosystem. There are two scenarios:
(i) If $r$'s are the same for all entries in the coefficient matrix,
then all the identical entries in $\mathbf{A}$ will be encrypted
to identical entries in $\mathbf{A}^{\prime}$. In other words, by
inspecting identical entries in $\mathbf{A}^{\prime}$, we can determine
whether entries in $\mathbf{A}$ are identical or not. However, in
CASO, since an entry in $\mathbf{A}^{\prime}$ is the linear combination
of entries in $\mathbf{A}$ and $\mathbf{K}$, the identical entries
in $\mathbf{A}^{\prime}$ would not indicate that the corresponding
entries in $\mathbf{A}$ are identical. Thus, in this case, CASO will
\emph{provide better security protection.} In this case, the end-user
needs to compute $n^{2}+1$ exponential operation. (ii) If a different
$r$ is used for each entry of the coefficient matrix, then the end-user
has to randomly select $n^{2}$ $r$'s, which is quite complex. Furthermore,
the end-user need to compute $2$ exponential operations for each
entry ($g^{a_{i,j}}$ and $r^{n}$). Therefore, altogether, the end-user
has to compute $2n^{2}$ exponential operations. In addition, due
to security requirement, $n$ has to be at least 1024 bits long. In
this case, $n^{2}$ would be 2048 bits. As an example, the size of
the outsourced coefficient matrix for 5000 variables would be around
6MB without data compression. While in scheme-1 and scheme-2 of our
proposed CASO, the transformation is applied in the column basis.
As a result, the order information of each column may be exposed.
In this sense, the scheme in \cite{wang2011icdcs} may provide better
protection than scheme-1 and scheme-2 regarding the coefficient matrix
$\mathbf{A}$ . However, in scheme-3, each entry in $\mathbf{A}$
is transformed to 
\[
a_{ij}^{\prime}=\sum_{r=j-\omega}^{j+\omega}a_{ir}k_{rj}.
\]
When $\omega>0$, since each $k_{rj}$ in $\mathbf{K}$ is randomly
chosen, the order information in each column will also be concealed.
Thus the scheme in \cite{wang2011icdcs} can provide comparable security
protection regrading the coefficient matrix $\mathbf{A}$ as scheme-3.
In scheme-4, the entries in $\mathbf{A}^{\prime}$ are further permuted.
As a result, there exist no explicit relation between the entry $a_{ij}$
in $\mathbf{A}$ and the corresponding entry $a_{ij}^{\prime}$ in
$\mathbf{A}^{\prime}$. However, one can know for sure that the entry
$t_{ij}^{\prime}$ in $\mathsf{Enc}(\mathbf{T})$ is encrypted from
the entry $t_{ij}$ in $\mathbf{T}$. Thus scheme-4 can provide better
protection of $\mathbf{A}$.

In terms of the solution vector $\mathbf{x}$, in \cite{wang2011icdcs},
the solution vector $\mathbf{x}$ is protected by adding a random
vector $\mathbf{r}$ as $\mathbf{y}=\mathbf{x}+\mathbf{r}$, while
in our scheme, we conceal $\mathbf{x}$ by the affine mapping $\mathbf{x}=\mathbf{Ky}+\mathbf{r}$.
Thus, CASO scheme can provide better security protection in this aspect.

\subsubsection{Convex Optimization}

In \cite{xu2013infocommini}, the authors proposed a verification
scheme for convex optimization problems. However, they did not give
any outsourcing scheme. Compare to \cite{xu2013infocommini}, in addition
to result verification, CASO also provides a secure outsourcing scheme.
Even in result verification, CASO outperforms it in terms of computational
complexity. 

The result verification of convex optimization is divided into three
categories: normal, infeasible and unbounded. The verification for
normal case forms the basis for other two cases. For the normal case,
the basic idea in \cite{xu2013infocommini} is to check the Karush-Kuhn-Tucker
(KKT) optimality condition. The end-user has to evaluate the original
functions as well as their differentials based on the optimal points
returned by the cloud. This verification process is much more expensive
since all the original functions are non-linear. In comparison, our
verification scheme requires only linear operations (e. g. multiplication
and addition) on the independent variables and the returned solution,
therefore, it must be more efficient.

\subsubsection{Summary}

We summarize the performance comparison of CASO with some existing
works in Table \ref{tab:Performance-Comparison}. We have shown that
in the case of outsourcing linear programming (LP) and system of linear
equations (LE), CASO outperforms the existing schemes in computational
complexity. In terms of security, all the schemes are secure in protecting
the original coefficient matrix. That is, given the disguised problem,
input and output, it is computational infeasible to recover the original
problem, input and output. CASO can also be applied to system of non-linear
equations (NLE) and convex optimization (COPT). This shows that CASO
possesses better applicability. Furthermore, compared to the existing
works, CASO also gives end-users the flexibility to choose the most
suitable outsourcing strategy on a cost-aware basis. That is the end
user can select the secret key $\mathbf{K}$ for the outsourcing scheme
based on its various security demands and computational resources. 

\begin{table*}
\def\arraystretch{1.5}

\caption{Performance Comparison \label{tab:Performance-Comparison}}

\centering{}	%
\begin{tabular}{|c|c|c|c|c|c|c|}
\hline 
\multirow{2}{*}{} & \multicolumn{4}{c|}{Applicability} & \multirow{2}{*}{Computational Complexity} & \multirow{2}{*}{Communication Overhead Index $\mathcal{I}_{c}$}\tabularnewline
\cline{2-5} 
 & LE & LP & NLE & COPT &  & \tabularnewline
\hline 
Our Scheme & $\surd$ & $\surd$ & $\surd$ & $\surd$ & $\mathcal{O}(n^{2})$ & 1\tabularnewline
\hline 
\cite{wang2011infocom} &  & $\surd$ &  &  & $\mathcal{O}(n^{2.376})$ & 1\tabularnewline
\hline 
\cite{wang2011icdcs} & $\surd$ &  &  &  & $\mathcal{O}(n^{3+\epsilon})$ & $\frac{1}{L}$\tabularnewline
\hline 
\cite{xu2013infocommini} &  &  &  & Only Verification & Not Applicable & Not Applicable\tabularnewline
\hline 
\end{tabular}
\end{table*}

\subsection{Numeric Results}

In this section, we measure the performance of CASO utilizing MATLAB.
The computation of both the end-user and the cloud server is simulated
using the same computer with an Intel Core 2 Due CPU running at 2.
53 GHz with 4GB RAM. We take outsourcing of the system of linear and
non-linear equations as examples. In the process of outsourcing, we
focus on the overhead of problem transformation, result recovery and
the performance gain that they can achieve by outsourcing problems
to the cloud. We denote the time for local computation in the outsourcing
process $\mathcal{T}_{e}$, the time cost without outsourcing $\mathcal{T}_{s}$,
and the performance gain $\mathcal{I}=\mathcal{T}_{s}/\mathcal{T}_{e}$. 

We first show the simulation results for outsourcing of system of
linear equations $\mathbf{Ax}=\mathbf{b}$, where $\mathbf{A}$ is
an $n\times n$ matrix. In complexity analysis, we show that the complexities
of scheme-1 and scheme-2 are in the same level while the complexity
for scheme-3 and scheme-4 are comparable. 

In scheme-3, when the bandwidth $W$ equals to $1$, it reduced to
scheme-1. Thus in our evaluation, we take scheme-3 as an example and
let $\mathbf{K}$ be a band matrix with bandwidth $W$ varying from
1 to 31. To investigate the impact of problem size on our proposed
scheme, we let $n$ vary from 1000 to 5000. The numeric results are
shown in Table \ref{Table:Evaluation-SLE}. First, we can learn from
the results that when the bandwidth of the banded matrix $\mathbf{K}$
becomes larger, the computational overhead at local side grows and
the performance gain decreases. This fact coincides with our analysis
of the trade off between complexity and security. Second, the performance
gain increases with the growth of the problem dimension $n$. This
is because our scheme requires the end-users to carry out simple operations
such as addition and multiplication. And this feature becomes more
obvious for the case of non-linear computation. 

\begin{table}
\def\arraystretch{1.5}

\caption{Performance Evaluation for System of Linear Equations \label{Table:Evaluation-SLE}}

\centering{}%
\begin{tabular}{|c|l|l|l|l|}
\hline 
\textbf{Dimension} & \textbf{Bandwidth} & \textbf{$\mathcal{T}_{e}$ (sec)} & \textbf{$\mathcal{T}_{s}$ (sec)} & \textbf{$\mathcal{I}$}\tabularnewline
\hline 
\multirow{4}{*}{$n=1000$} & $W=1$ & $0.0265$ & $0.2356$ & $8.9$\tabularnewline
\cline{2-5} 
 & $W=7$ & $0.0265$ & $0.2402$ & $9.1$\tabularnewline
\cline{2-5} 
 & $W=15$ & $0.0546$ & $0.2356$ & $4.3$\tabularnewline
\cline{2-5} 
 & $W=31$ & $0.0858$ & $0.2387$ & $2.8$\tabularnewline
\hline 
\multirow{4}{*}{$n=2000$} & $W=1$ & $0.0593$ & $1.3962$ & $23.6$\tabularnewline
\cline{2-5} 
 & $W=7$ & $0.0936$ & $1.4071$ & $15.0$\tabularnewline
\cline{2-5} 
 & $W=15$ & $0.1248$ & $1.3853$ & $11.1$\tabularnewline
\cline{2-5} 
 & $W=31$ & $0.1950$ & $1.3494$ & $6.9$\tabularnewline
\hline 
\multirow{4}{*}{$n=3000$} & $W=1$ & $0.1170$ & $3.9234$ & $33.5$\tabularnewline
\cline{2-5} 
 & $W=7$ & $0.1856$ & $3.9281$ & $21.2$\tabularnewline
\cline{2-5} 
 & $W=15$ & $0.3058$ & $3.8844$ & $12.7$\tabularnewline
\cline{2-5} 
 & $W=31$ & $0.4867$ & $3.8766$ & $8.0$\tabularnewline
\hline 
\multirow{4}{*}{$n=4000$} & $W=1$ & $0.2184$ & $8.5832$ & $39.3$\tabularnewline
\cline{2-5} 
 & $W=7$ & $0.3416$ & $8.6924$ & $25.4$\tabularnewline
\cline{2-5} 
 & $W=15$ & $0.7129$ & $8.6565$ & $12.1$\tabularnewline
\cline{2-5} 
 & $W=31$ & $1.0171$ & $8.6768$ & $8.5$\tabularnewline
\hline 
\multirow{4}{*}{$n=5000$} & $W=1$ & $0.3260$ & $15.8138$ & $48.5$\tabularnewline
\cline{2-5} 
 & $W=7$ & $0.5288$ & $15.9839$ & $30.2$\tabularnewline
\cline{2-5} 
 & $W=15$ & $1.2683$ & $15.8793$ & $12.5$\tabularnewline
\cline{2-5} 
 & $W=31$ & $1.8174$ & $15.9698$ & $8.8$\tabularnewline
\hline 
\end{tabular}
\end{table}

Then we show the performance of our proposed scheme for system of
non-linear equations. We assumes that the non-linear system is composed
of polynomials on ten variables and let the number of independent
terms $N$ vary from 1000 to 5000. Also for the same reason, we deploy
band matrix as the key matrix and let the bandwidth $W$ vary from
1 to 3. The simulation result is shown in Table \ref{Table::Performance-Non-linear}.
For system of non-linear equations, the performance gain is larger
than its linear counterpart. This is because CASO requires only linear
operations (e. g. multiplication and addition) in the local environment.
Similar to that of the system of linear equations, the results clearly
show that there exists a trade-off between the computational complexity
and security. 

\begin{table}
\def\arraystretch{1.5}

\caption{Performance Evaluation for System of Non-linear Equations \label{Table::Performance-Non-linear}}

\centering{}%
\begin{tabular}{|c|c|c|c|c|}
\hline 
\textbf{Dimension} & \textbf{Bandwidth} & \textbf{$\mathcal{T}_{e}$ (sec)} & \textbf{$\mathcal{T}_{s}$(sec)} & \textbf{$\mathcal{I}$}\tabularnewline
\hline 
\multirow{3}{*}{$N=1000$} & $W=1$ & $1.6800$ & $26.2$ & $15.6$\tabularnewline
\cline{2-5} 
 & $W=2$ & $2.4500$ & $27.1$ & $11.1$\tabularnewline
\cline{2-5} 
 & $W=3$ & $3.0500$ & $26.2$ & $8.6$\tabularnewline
\hline 
\multirow{3}{*}{$N=2000$} & $W=1$ & $3.1500$ & $118.2$ & $37.5$\tabularnewline
\cline{2-5} 
 & $W=2$ & $5.1200$ & $118.8$ & $23.2$\tabularnewline
\cline{2-5} 
 & $W=3$ & $6.3900$ & $117.1$ & $18.3$\tabularnewline
\hline 
\multirow{3}{*}{$N=3000$} & $W=1$ & $5.1300$ & $330.8$ & $64.5$\tabularnewline
\cline{2-5} 
 & $W=2$ & $7.7100$ & $313.0$ & $40.6$\tabularnewline
\cline{2-5} 
 & $W=3$ & $9.7500$ & $320.6$ & $32.9$\tabularnewline
\hline 
\multirow{3}{*}{$N=4000$} & $W=1$ & $7.1600$ & $712.9$ & $99.6$\tabularnewline
\cline{2-5} 
 & $W=2$ & $12.3800$ & $713.1$ & $57.6$\tabularnewline
\cline{2-5} 
 & $W=3$ & $13.9000$ & $711.4$ & $51.2$\tabularnewline
\hline 
\multirow{3}{*}{$N=5000$} & $W=1$ & $9.3700$ & $1187.2$ & $126.7$\tabularnewline
\cline{2-5} 
 & $W=2$ & $16.0000$ & $1190.1$ & $74.4$\tabularnewline
\cline{2-5} 
 & $W=3$ & $20.6700$ & $1191.1$ & $57.6$\tabularnewline
\hline 
\end{tabular}
\end{table}

\section{Conclusion\label{sec:Conclusion}}

In this paper, we proposed a cost-aware secure outsourcing scheme
(CASO) for general computational problems. We demonstrated that CASO
can be utilized for secure outsourcing of various computational problems,
such as system of equations, linear programming and convex optimizations.
Our scheme also provides mechanisms for the end-users to verify results
received from the cloud. We provided security analysis on our proposed
scheme on a cost-aware basis. In particular, we proved that CASO is
secure in protecting the coefficient matrix of the outsourced problem
and can partly conceal the side information. Our analysis shows that
CASO can limit the computational overhead at the local side to $\mathcal{O}(n^{2})$.
Since CASO is executed off-line, the communication overhead is in
the same level as that of outsourcing the original problem itself.
We also compared CASO with several existing schemes and showed that
CASO is more efficient and has a wider applicability.

\end{document}